\documentstyle[prb,preprint,aps]{revtex}
\input epsf
\tighten
\begin{document}
\draft
\title{Fractional quantum Hall effect in higher Landau levels}
\author{Lotfi Belkhir\cite{lb} and J. K. Jain\cite{jkj}}
\address{Department of Physics, S.U.N.Y at Stony Brook, Stony Brook,
N.Y, 11794-3800}
\date{\today}
\maketitle
\begin{abstract}
\widetext
\advance\leftskip by 57pt
\advance\rightskip by 57pt

We investigate, using finite size numerical calculations, the spin-polarized
fractional
quantum Hall effect (FQHE) in the first excited Landau level (LL). We find
evidence for the existence of an incompressible state at
$\nu = \frac{7}{3} = 2+\frac{1}{3}$, but not at $\nu = 2+\frac{2}{5}$.
Surprisingly, the 7/3 state is found to be strongest at a finite
thickness. The structure of the low-lying excited states is found to
be markedly different from that in the lowest LL.
This study also rules out FQHE at a large number of odd-denominator
fractions in the lowest LL.

\end{abstract}
\pacs{PACS numbers: 7340Hm,7320Dx}

A good microscopic understanding of the physics of the
fractional quantum Hall effect (FQHE) in the lowest Landau level (LL)
is provided by the composite fermion theory \cite{jain1}.
It is straightforwardly explained as the integer quantum Hall effect (IQHE)
of a novel kind of particles, namely composite fermions, composed of
electrons bound to an even number of vortices.
The recent experimental observation of composite
fermions by several groups\cite{Stormer} has given
further support to this picture. Understanding of FQHE in the first
excited LL is not so satisfactory, however.

Several experiments have shown clear evidence of FQHE in
the first excited  LL as well. Taking account of both
spin-subbands of the lowest LL (LLL), the first excited LL
corresponds to the experimental filling factor range $4>\nu>2$.
The first observed fraction in this range
was $\nu = \frac{5}{2} = 2+\frac{1}{2}$\cite{willet,eisen1}.  The occurrence
of an incompressible  state at $\nu=\frac{5}{2}$ is quite peculiar for several
reasons. One, that  there is no FQHE at $\nu=\frac{1}{2}$, its counterpart
in the LLL. In fact,
recent experiments at $\nu=\frac{1}{2}$\cite{Stormer} showed clear
evidence of a Fermi surface of spin-polarized
composite fermions at this filling factor.
Second, $\nu = \frac{5}{2}$ is so far the only even denominator fraction
observed in single layer systems. Finally, exact
finite size calculations, which involve only Coulomb interaction between
electrons within one LL, do not seem to find any incompressible state at
this filling factor.
Several attempts have been made to provide a theoretical understanding
of this state.
Haldane and Rezayi\cite{hr} proposed a spin-singlet wave function for a
$\nu=\frac{5}{2}$ FQHE, which is valid for a hollow-core model
interaction. The present authors proposed a different spin-singlet
wave function,
which is valid for a short-range, {\sl hard-core} model interaction,
and possesses the composite fermion structure \cite {bj}.
Neither of these wave functions is a good representation of the Coulomb ground
state, though. It is possible that inclusion of higher Landau level
mixing might make one of these relevant to 5/2 FQHE.
The present work only deals with fully polarized states.

Another fraction in higher LL at which FQHE has been observed is
$\nu=\frac{7}{3}=2+\frac{1}{3}$.
Laughlin's successful wave function at 1/3 \cite {Laughlin}
has been generalized to 7/3 \cite {MD}, but it does not provide a
very satisfactory description of the 7/3 state, consistent with
a relatively weak FQHE at this fraction.  Our calculations confirm
this result for strictly two-dimensional (2D) systems.
However, we find the surprising result that
for wider quantum wells, the exact ground state
becomes closer to the Laughlin state, and the FQHE at 7/3 becomes
stronger; the energy gap acquires its maximum value when the
thickeness is roughly equal to twice the magnetic length.
This is in contrast to the situation in the lowest LL, where
the FQHE is in general the strongest at zero thickness.

We model the finite thickeness effects, in the square well configuration,
by accounting for the spread of the one-electron wave function along the
z-direction. (We have also studied the triangular heterojunction
confinement; the results are similar.)
We restrict ourselves to the lowest subband;
this approximation is not valid for large thicknesses, when the
subband spacing is small, and interband transitions become important.
In the square well geometry, the (unnormalized) z-component of the
one-electron wave function
in the lowest band is given by $\xi(z)=\cos(z/d)$,
where $d$ is the thickeness of the well.  The distances will be
expressed in units
of the magnetic length, $\lambda$. We make use of Haldane's
pseudopotentials $V_m$\cite{Hrev}, which are the energies of pairs of
electrons with relative angular momentum $m$:

\begin{equation}\label{Vm}
V_{m} = \int_{0}^{\infty}~qdq~\tilde{V}(q)\left[L_n(\frac{q^2}{2})
\right]^2~L_m(q^2)~e^{-q^2}.
\end{equation}
where $n$ denotes the LL index,  and $~\tilde{V}(q)$ the electron-electron
interaction in momentum space given by

\begin{equation}\label{Vq}
\tilde{V}(q) = \frac{q^2}{\kappa~q}\int~dz_1\int~dz_2
\left|\Psi(z_1)\right|^2\left|\Psi(z_2)\right|^2~e^{-|z_1-z_2|q}
\end{equation}
It is convenient to write the Coulomb interaction in the form
\begin{equation}\label{Vq2}
\tilde{V}(q) = \frac{q^2}{\kappa~q}~F(q)
\end{equation}
where the net effect of finite thickeness is now summed up in the form
factor $F(q)$. In the case of a square well potential, it is given
by\cite{DasSarma}
\begin{equation}\label{Fq}
F(q) = \frac{1}{x^2+4\pi^2}\left[3x + \frac{8\pi^2}{x} -
\frac{32\pi^4(1-\exp^{-x})}{x^2(x^2+4\pi^2)}\right]; x = qd
\end{equation}

Our calculations are performed in the spherical geometry\cite {Hrev}, in which
electrons move on the surface of a sphere under the influence of a
radial magnetic field. The flux through the surface, $N_{\phi}$, measured
in unit of the flux quantum $\phi_{0}=hc/e$, must be an integer.
The degeneracy of the LL is $N_{\phi}+1$ and
increases by 2 for each successive LL. It is well
known at what values of $N_{\phi}$ various
incompressible states (e.g. 1/3 and 2/5) occur in the LLL
for finite systems \cite {Hrev,Dev}. It is assumed that in the first
excited LL, 7/3 (12/5) occurs at that value of magnetic field
for which the degeneracy of the first excited LL is equal to the
degeneracy of the LLL at 1/3 (2/5).
In all our calculations, we use the thermodynamic values of $V_{m}$.
Only electrons in a given
LL are considered (i.e., LL mixing is neglected), and it
is assumed that they are fully polarized.
While comparing the LLL state with the corresponding higher LL state,
we use the standard prescription \cite {MD} in which the LL index
is changed before computing the overlaps.  As usual, the eigenstates
are labeled by their total orbital angular momentum $L$.

Our calculated
energy spectra of 6 and 8 electrons at filling fractions of $\frac{7}{3}$
and $\frac{12}{5}$, respectively, are shown in Fig. \ref{spec}.
The two top figures correspond to filling factors 1/3 and 2/5 (in the
lowest $n = 0$ LL) at zero thickness, while the rest of the figures
correspond to 7/3 (left column) and 12/5 (right column)
for different values of thickeness.
The spectra in the LLL have been explained in great detail using the
composite fermion theory \cite {Dev}.
At $\nu=\frac{7}{3}$, the ground state is at $L=0$, and
appears to have a gap, which is similar to the situation at 1/3.
The collective mode branch, however, seems different from
that at 1/3; it extends to smaller $L$, and no clear roton minimum
may be identified. The spectrum at $\nu=\frac{12}{5}$ looks completely
different from that at $\nu=2/5$; even the ground state is not uniform.
To learn more about the nature of the low lying states,
we calculate their overlap with the corresponding states in the LLL,
shown in figure \ref{spec}.  While the overlaps are poor for strictly 2D
layers ($d=0$), they increase appreciably as the thickness of the layers
is increased. For
$\nu=\frac{12}{5}$ a level crossing transition occurs to a uniform
($L=0$) ground state at around
around $d=2\lambda$, but there is no clear energy gap even at large widths.

We now study bigger systems, and
focus on the nature of the incompressible ground state at
$\nu=\frac{7}{3}$ for an 8 electron system  and at $\nu=\frac{12}{5}$ for
a 10 electron system. The sizes of the Hilbert spaces in the $L_{z}=0$
sector are 8512 and 16660, respectively, and a Lanczos
algorithm is used for obtaining the ground state.
Figure \ref{over8}-a shows the overlap of the $\nu=\frac{7}{3}$ ground state
with the LLL 1/3 ground state.
At zero thickeness, the overlap is slightly
below 0.8. As the sample thickness is increased, the overlap increases
to 0.93 at $d=4$. Notice that the overlap at $d=4$ is slightly larger for
$N=8$ than for $N=6$ shown in figure \ref{spec}, suggesting
that the large overlap is {\sl not} a finite size effect, and that the
$\nu=\frac{7}{3}$ state is well described by the Laughlin
wave function. The picture is quite different at $\nu=\frac{12}{5}$.
Figure \ref{over8}-b shows the overlap of the ground state at
$\nu=\frac{12}{5}$ for a 10
electron system with the ground state at 2/5.
The overlap is extremely poor for all values of
the thickeness, indicating that the relatively large overlap obtained
for $N=6$ was probably a finite size effect, and that there is, in
fact, no FQHE at 12/5 in the thermodynamic limit.

Of particular interest in the  FQHE is the value of the energy gap of
the incompressible states, since it is an experimentally accessible quantity.
In particular, it was found experimentally
\cite{Goldman} that the gap of the 1/3
state decreases rapidly with increasing thickness.
Song He {\sl et. al}\cite{He} investigated this theoretically, and
found that the gap decreases because both the Coulomb interaction and
the incompressible state become weaker as the thickness is increased,
the latter manifested in the decreasing overlap of the ground state
with the Laughlin state.
In the first excited LL the picture is quite different, since
the two effects compete: while the Coulomb interaction gets
weaker, the overlap of the ground state
with the Laughlin wave function {\em increases} as a function of thickness.
Figure \ref{Gap8_13} shows the energy gap for the 8 electron system
as a function of thickeness \cite{gap}; it attains its maximum value at $d
\approx 2$, which is about  40\% larger than the $d=0$ gap. It
would be quite interesting to see experimentally this
enhancement of the energy gap.

An insight into the {\em qualitative} difference between the two
LL's can be gained in terms
of dimensionless scaled pseudopotentials\cite{He}, defined as
\begin{equation}\label{fm}
f_m = (V_3 - V_m)/(V_1 - V_3)
\end{equation}
They explicitly satisfy the property that they are invariant under a
constant shift of $V_m \rightarrow V_m+C$, which neither alters the
eigenstates nor the eigenenergies (measured relative to the ground
state energy). Furthermore, $f_1=-1$ and $f_3=0$ are independent of
$V_{m}$.  The Laughlin 1/3 state is the exact ground state for a hard-core
model in which all $V_{m}$'s are zero except $V_{1}$; i.e., when
$f_{1}=-1$ and all other $f_{m}$'s are zero.
So the nonzero values of $f_m$ characterize the deviation
from the hard-core model, and thus from the Laughlin state.
At $\nu=1/3$, the scaled pseudopotentials
increase monotonically with thickness \cite {He}, as shown in figure
\ref{fmfig}-a, so
that at large enough thicknesses,
the deviation is so large that the Laughlin state is destroyed.

Figure \ref{fmfig}-b shows the thickness dependence of the
${f_m}$ in the
$n=1$ LL. For $m\geq 7$, ${f_m}$ do  not increase
monotonically,
but rather decrease first, mark a soft minimum, and then increase as
a function of thickness. Moreover, $f_5$, which has the dominant
effect,
decreases for all values of thickness. The deviation from the
hard-core
model is therefore reduced by finite thickness in the $n=1$ LL, which
helps understand why the overlap of the $\nu=\frac{7}{3}$ state with
the Laughlin state increases, and reaches a plateau at large thickness.

Since the composite fermion theory relates QHE in higher LL's to FQHE
in the lower LL's, our study also puts strong constraints on
which fractions may be observed in
the LLL. The principal observed fractions in the LLL correspond to
the IQHE of composite fermions; e.g., $\nu=n/(2n+1)$ FQHE of electrons
corresponds to
$\nu^*=n$ IQHE of composite fermions.  The {\em FQHE} of composite
fermions will lead to new fractions for electrons.
Let us assume very large Zeeman
energy, i.e., spinless electrons, so that a filling factor $\nu$ in
the first excited LL corresponds to an overall filling factor of
$1+\nu$. Then, the prominent FQHE in the first excited LL is expected
to occur at are
$\nu=1+n/(2n+1)$ and $\nu=2-n/(2n+1)$. FQHE of composite fermions at
these filling factors corresponds to FQHE of electrons in the LLL at
$\nu=(3n+1)/(8n+3)$ and $\nu=(3n+2)/(8n+5)$, respectively.
Our study shows that no FQHE can occur for $n\geq 2$. In other words,
the only states between 1/3 and 2/5 where FQHE may occur are
4/11 and 5/13 (which correspond to $n=1$).
Since 1/3 is already quite weak in the first excited LL, it is likely
that {\em no} FQHE is observed in third and higher LL's, which leads to the
prediction that no FQHE states other than $n/(2n+1)$ are possible in
the filling factor range $2/5\leq \nu \leq 1/2$.
These predictions are generally consistent with experiments; there is
some evidence for 4/11, but no other non-$n/(2n+1)$
fractions have been observed in this range.  These considerations can
be easily generalized to other regions of filling factors.

In conclusion, we have found that fractional quantum Hall effect occurs
only at $\nu=\frac{1}{3}$  and $\nu=\frac{2}{3}$ in the first excited LL,
and that it is strongest at a finite thickness.
We thank Professor A. MacDonald for fruitful comments on this work.
This work was supported by the NSF under Grant no. DMR9318739.

\newpage
\begin{figure}
\caption{ Energy spectra of the low lying states for
6 (left column) and 8 (right column) electrons.
The two top spectra are in the lowest
($n=0$) LL, and the rest in the first excited
($n=1$) LL. The filling factor and the width are shown above each
spectrum. The energies are given in units of $e^2/\lambda$.}
\label{spec}
\end{figure}

\begin{figure}
\caption{
(a) Overlap of the ground state wave function at $\nu=\frac{7}{3}$ (1/3
in the first excited LL), with the ground state in
lowest LL for 8 electrons.
(a) Overlap of the ground state wave function at $\nu=\frac{12}{5}$ (2/5
in the first excited LL), with the ground state in lowest LL for 10 electrons.
}
\label{over8}
\end{figure}

\begin{figure}
\caption{
Energy gap of $\nu=\frac{7}{3}$ as a function of thickness.
{}~~~~~~~~~~~~~~~~~~~~~~~~~~~~~~~~~~~~~~~~~~~~~~~~~~~~~~~~~~~~~~~~~~~~~
}
\label{Gap8_13}
\end{figure}

\begin{figure}
\caption{
The scaled pseudopotentials, defined in the text, as a function of thickness
for (a) the lowest and (b) the first excited LL.
}
\label{fmfig}
\end{figure}


\begin{references}
\bibitem{gap} The gap is computed from the size of the cusp at 7/3.
\end{references}
\end{document}